\documentclass[twocolumn,pre,aps,showpacs,floatfix]{revtex4}

\usepackage{graphicx,amssymb,amsmath}

\newcommand{\cbm}{C_B^\mathrm{max}}
\newcommand{\ibm}{c_B^\mathrm{max}}
\newcommand{\il}{l^{-1}}
\newcommand{\aS}{\langle S\rangle}

\begin{document}

\title{Edge overload breakdown in evolving networks}

\author{Petter \surname{Holme}}
\email{holme@tp.umu.se}
\affiliation{Department of Theoretical Physics, Ume{\aa} University,
  901 87 Ume{\aa}, Sweden}

\begin{abstract}
  We investigate growing networks based on Barab\'{a}si and Albert's
  algorithm for generating scale-free networks, but with edges sensitive
  to overload breakdown. The load is defined through edge betweenness
  centrality. We focus on the situation where the average number of
  connections per vertex is, as the number of vertices, linearly
  increasing in time. After an initial stage of growth, the network
  undergoes avalanching breakdowns to a fragmented state from which it
  never recovers. This breakdown is much less violent if the growth is
  by random rather than preferential attachment (as defines the
  Barab\'{a}si and Albert model). We briefly discuss the case where
  the average number of connections per vertex is constant. In this
  case no breakdown avalanches occur. Implications to the growth of
  real-world communication networks are discussed.
\end{abstract}

\pacs{89.75.Fb, 89.75.Hc}
%89.75.Fb Structures and organization in complex systems
%89.75.Hc Networks and genealogical trees

\maketitle

\section{Introduction}
Large sparse networks are the underlying structure for transportation
or communication systems, both man made (like computer
networks~\cite{AJB,otherCN} or power grids~\cite{PG}) or natural (like
neural networks~\cite{NN} or biochemical networks~\cite{BioC}). These
networks displays both randomness and some self-induced structure
influencing the flow of transport and robustness against congestion or
breakdown in the network. One of the most conspicuous structures among
real-world communication networks is a highly skewed distribution of
the degree (the number of neighbors of a
vertex)~\cite{AJB,otherCN,otherSF}.

Avalanching breakdown in networks where the edges or vertices are
sensitive to overload is a serious threat for real-world networks. A
recent example being the black-out of 11 US states and two Canadian
provinces the 10th August 1996~\cite{blackout}. Recently the overload
breakdown problem for vertices in growing networks with an emerging
power-law degree distribution has been studied~\cite{EGO2}. In the
present paper we investigate the overload breakdown problem when edges
(rather than vertices) are sensitive to overloading. We use the
standard model for such networks---the Barab\'{a}si-Albert (BA)
model~\cite{BA1,BA2}, but with a maximum load capacity assigned to
each  edge. The load is defined by means of the betweenness
centrality---a centrality measure for communication and transport flow
in a network~\cite{BC}. The procedure enables us to study overload
breakdown triggered by the redistribution (and increase) of load in a
growing network. This is in contrast to earlier models of cascading
breakdown phenomena, all dealing with vertex breakdown, that has taken
a fixed network as their starting point~\cite{WATTS,MGP}.

\section{Definitions}
We represent networks as undirected and unweighted graphs $G=(V,E)$
where $V$ is the set of vertices, and $E$ is the set of unweighted
edges (unordered pairs of vertices). Multiple edges between the same
pair of vertices are not allowed.

\subsection{The Barab\'{a}si-Albert model of scale-free networks}
The standard model for evolving networks with an emerging power-law
degree distribution is the Barab\'{a}si-Albert model. In this model,
starting from $m_0$ vertices and no edges, one vertex with $m$ edges
is attached iteratively. The crucial ingredient is a biased selection
of what vertex to attach to, the so called ``preferential attachment:''
In the process of adding edges, the probability $P_u$ for a new vertex 
$v$ to be attached to $u$ is given by~\cite{BA3}
\begin{equation}
  P_u = \frac{k_u+1}{\sum_{w\in V}(k_w+1)}\label{paeq} ~,
\end{equation}
where $k_u$ is the degree of the vertex $u$. To understand the effect
of preferential attachment, we will also investigate networks grown
with a unbiased random attachment of vertices. Without the
preferential attachment the networks are known to have an exponential
tail of the degree distribution~\cite{BA2}. The time $t$ is measured
as the total number of added edges, which is different by factor $m$
from Refs.~\cite{BA1,BA2} where $t$ is defined
as the number of added vertices.

It should be noted that in very large communication networks, such as
the Internet, the users can process information about only a subset of
the whole network. How this affects the dynamics of network formation
is investigated in Ref.~\cite{mossa}. In the present work we neglect
such effects and assume linear preferential attachment.

\subsection{Load and capacity}\label{sect:load}
To assess the load on the vertices of a communication network, or any
network where contact between two vertices are established through
a path in the network, a common choice is the betweenness
centrality~\cite{BC} which often is seen as a vertex quantity but
has a natural extension to edges $e\in E$~\cite{GN}:
\begin{equation} \label{eq:CB}
  C_B(e)=\sum_{v\in V}\sum_{w\in V\setminus\{v\}}\frac{\sigma_{vw}(e)}
  {\sigma_{vw}}~,
\end{equation}
where $\sigma_{vw}(e)$ is the number of geodesics between
$v$ and $w$ that contains $e$, and $\sigma_{vw}$ is the total number
of geodesics between $v$ and $w$. $C_B(e)$ is thus the number of
geodesics between pairs of vertices passing $e$; if more than one
geodesics exists between $v$ and $w$ the fraction of vertices
containing $e$ contributes to $e$'s betweenness.

In Ref.~\cite{EGO2} (see also Ref.~\cite{KAHNG}) the use of
betweenness centrality as a load measure is given thorough
motivations. These arguments are readily
generalized to the case of edges sensitive
to overloading: Suppose that $\Lambda$ is the set of pairs of vertices
with  established communications through shortest paths at a given
instant~\cite{foot}. Then let $\lambda(e)$ denote the load of
$e\in E$ defined as the number of geodesics that contains $e$. Then we
assume the effective load to be the average
\begin{equation}
  \langle\lambda(e)\rangle_\Omega =\frac{1}{|\Omega|}
  \sum_{\Lambda\in\Omega}\lambda(e) ,\label{eq:lambda}
\end{equation}
where $\Omega$ is an ensemble of $\Lambda$. To proceed, we restrict
$\Omega$ according to:
\begin{equation}
  \Omega=\{\Lambda:|\Lambda|=AN(N-1)\}, \label{eq:load_eca}
\end{equation}
where $A$ is constant with respect to
$N$. This is to be interpreted that an element of $\Omega$
is a set of $AN(N-1)$ pairs of distinct vertices
chosen uniformly at random, and thus corresponds to the case
where the number of established communication routes ending at a
specific vertex in average increases with $N$. This case can for
example be expected in the early days of the Internet where the
launches of new sites made the users browse a larger average number of
sites. The case where the users at average connects to an
$N$-independent number of others is discussed in
Appendix~\ref{sec:app}. The largest approximation, when using the
betweenness as a load measure, is probably that routing protocols of
e.g.\ the Internet has implicitly implemented load
balancing~\cite{foot,OSPF,huitema}.

To introduce overloading to the dynamics we assign a capacity, or
maximum value, $\lambda^\mathrm{max}(e)$ to the load, the same for
each edge, and say that the edge $e$ is overloaded if
$\lambda^\mathrm{max}(e)< \langle\lambda(e)\rangle_\Omega$. From the
definition of $\Omega$ we can see that our situation corresponds to
having a maximum capacity on the betweenness centrality of the edges
so that an edge is overloaded if $C_B(e)>\cbm$ (where $\cbm$ is
constant). If an edge is overloaded it is simply removed from the
graph, and the betweenness recalculated, if then another edge becomes
overloaded it is removed, and so on. If more than one edge is
overloaded at a time we choose the one to remove randomly. Multiple
breakdowns during one time step defines a ``breakdown avalanche.''

\subsection{Quantities for measuring network functionality}
To measure the network functionality we consider three quantities---the
number of edges $L$, inverse geodesic length $\il$, and the size of
the largest connected subgraph $S$: For the original BA model the
number of edges increases linearly as $L(t)=t$ (i.e.\ one edge is
added in unit time). But if an overload breakdown occurs in the system
$L$ decreases, making it a suitable simplest-possible-measure of the
network functionality. In a functional network a large portion of the
vertices should have the possibility to connect to each other. In
percolation and attack vulnerability studies of random networks one
often uses $S$ to define the system as `percolated' (or functioning),
when the size of the largest connected subgraph $S$ scales as
$N$~\cite{AJB,ATTPER}. One of the characteristic features of the BA model
networks, as well as many real-world communication networks, is a less
than algebraically increasing average geodesic length $l$. As the
average geodesic length is infinite when the network is disconnected
(as could be the case when an overload breakdown has occurred) we
study the average inverse geodesic length~\cite{NEWMAN}:
\begin{equation}
  \il\equiv\left\langle\frac{1}{d(v,w)}\right\rangle\equiv 
  \frac{1}{N(N-1)}\sum_{v\in V}\sum_{w\in V\setminus\{v\}}
  \frac{1}{d(v,w)}~,
  \label{eq:il}
\end{equation}
which has a finite value even for the disconnected graph
if one defines $1/d(v,w)\equiv 0$ in the case that no path connects
$v$ and $w$. To monitor the fragmentation of the network we will also
measure the number of connected subgraphs $n$.

\begin{figure}
  \centering{\resizebox*{8.4cm}{!}{\includegraphics{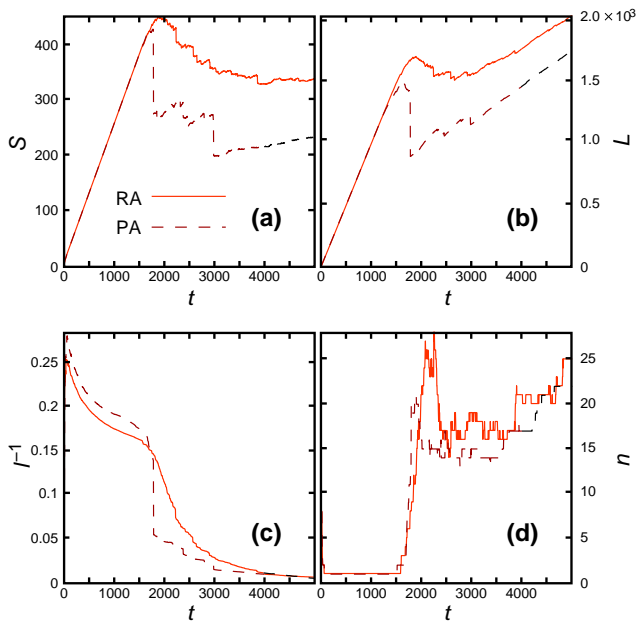}}}
  \caption{The time evolution of $S$ (a), $L$ (b), $\il$ (c), and $n$
    (d) for a typical run with $\cbm= 500$ and $m_0/2=m=4$. Dashed
    lines represent the network grown with preferential attachment
    (PA), solid grey lines denotes curves for the runs with an
    unbiased random attachment (RA).
    }
  \label{ECAcase}
\end{figure}

\section{Simulation results}
For relative small $m$, typical runs are exemplified in
Fig.~\ref{ECAcase}. For both random and preferential attachment $\aS$
reaches a critical time where after the network starts to break down,
eventually $\aS$ reaches a steady state value. The breakdown develops
differently in the two cases: For the random attachment the
breakdown is relatively slow and the steady state value is high
compared to the preferential attachment case where large successive
avalanches fragments the network. The other two quantities reflects
the same behavior: While the initial vertices gets joined into the
network $\il$ increases to an early maximum. After the decrease
corresponding to the increase of $l$, $\il$ decreases
rapidly when the network becomes fragmented. $L$ shows the jagged
shape, as expected  correlated with that of $\aS$. As seen in
Fig.~\ref{ECAcase}(a) and (b), the discontinuity in $L$ (in the
preferential attachment case), is less pronounced than
that in $\aS$, so a small number of overloaded edges can be
enough to cause large decrease in $\aS$. The reason for this behavior
is that bridges (single edges interconnecting connected subgraphs)
have a high betweenness and thus are prone to overloading. The number
of connected subgraphs behaves qualitatively the same for random and
preferential attachment. For other runs of the algorithm the breakdown
can qualitatively be described as above. The averaged quantities varies 
relatively little, for example the peak-time for $\aS$ has a
standard deviation of $\sim 3$\%.

The corresponding overload case for vertices studied in Ref.~\cite{EGO2}
shows a similar time development with an period of incipient
scale-freeness, an intermediate regime of breakdown and recovery
(although the period of recovery is not as large for edges as for
vertices), and a final breakdown to a large-$t$ state of disconnected
clusters. One major difference between overload breakdown for vertices
and edges is that the difference between random and preferential
attachment is larger for edge overloading---edge robustness benefits
more than vertex robustness from the geometry arising from random
attachment.

\begin{figure}
  \centering{\resizebox*{8cm}{!}{\includegraphics{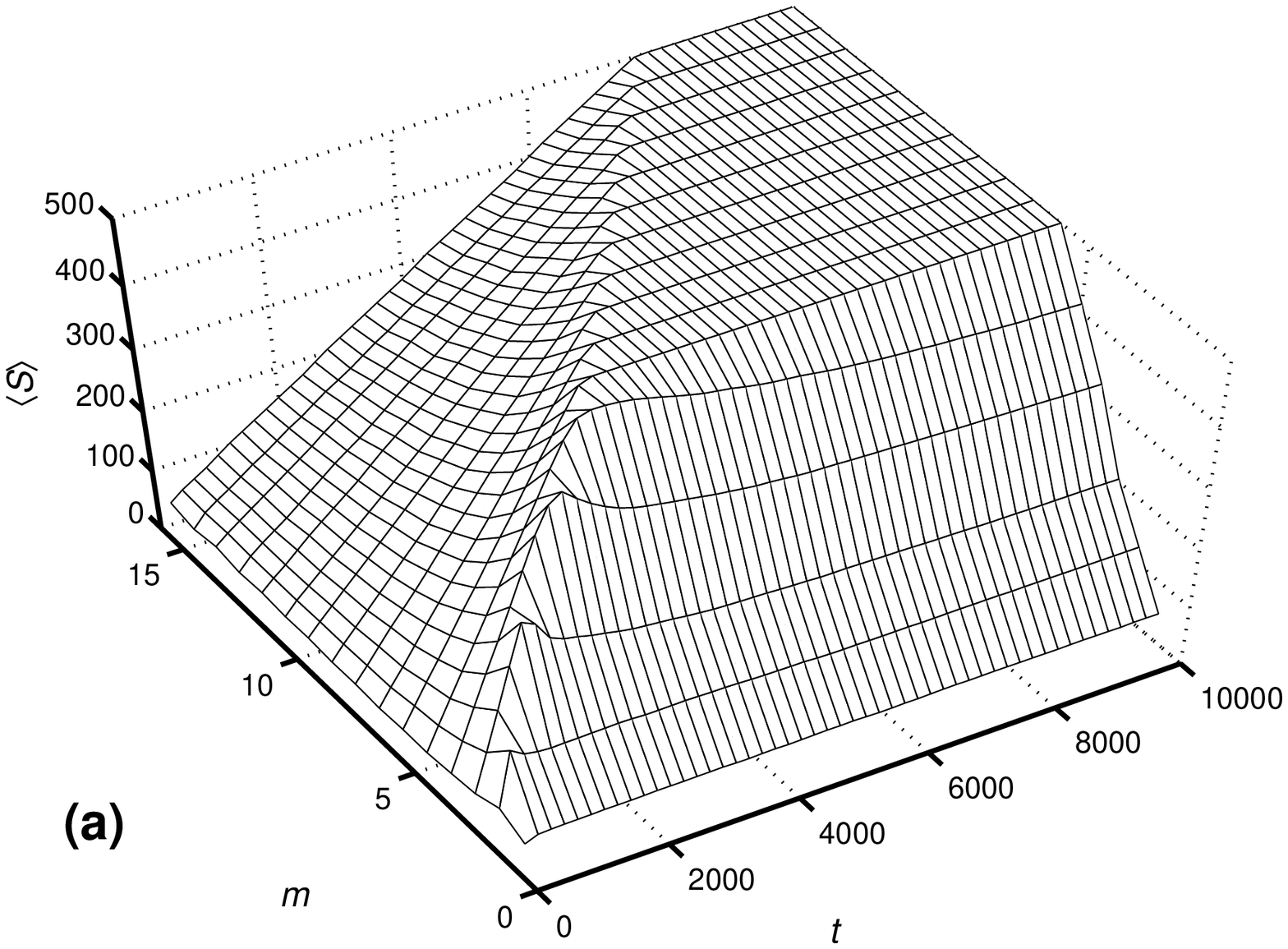}}}
  \centering{\resizebox*{8cm}{!}{\includegraphics{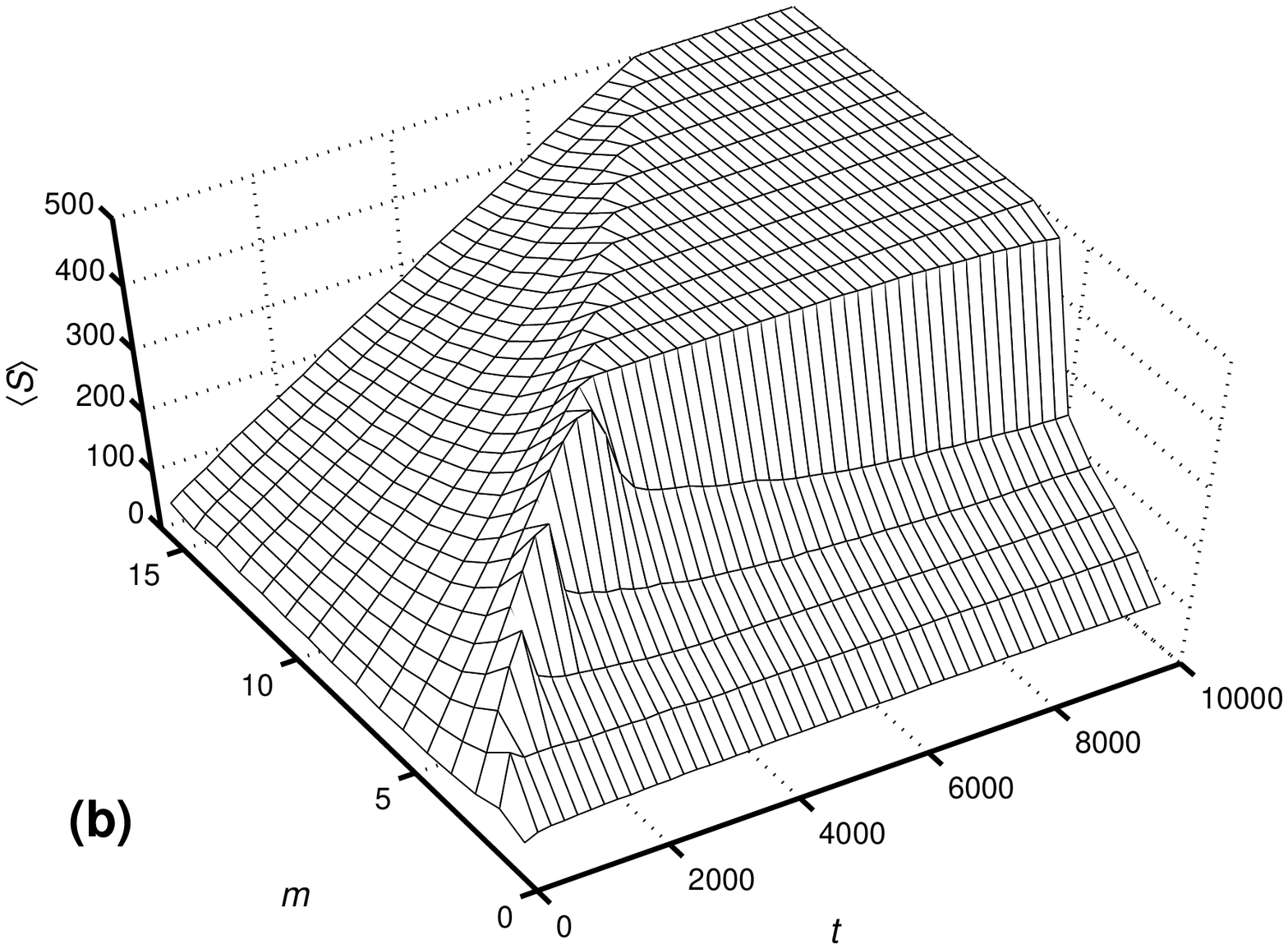}}}
  \caption{The $m$ dependence of the time development of $\aS$---the average
    size of the largest connected subgraph with $\cbm=500$,
    and $m_0=2m$, for (a) random attachment and (b) preferential
    attachment.}
  \label{lECA}
\end{figure}

Next we investigate the $m$-dependence. As seen
in Fig.~\ref{lECA} the system becomes more and more robust when $m$
increases. This is of course expected since with a higher average degree
more edges shares the load, so the maximal load can be expected to decrease.
For high enough $m$ there are no avalanches, the largest connected component
remains of the same size $S=\cbm+1$. When $S=\cbm+1$ the next edge attaching
a new vertex will have $C_B(e)=\cbm+1$, and thus be overloaded. In most
cases this will lead to removal of the newly added edge---otherwise
another edge has to be overloaded at the same time, which is
decreasingly likely with increasing $m$. In Fig.~\ref{lECA}(b) we can
see one exception to this interpretation at $m=6$: Here $\aS$ reaches
$\cbm+1$ but starts to decay slowly at around $t=8000$. As mentioned,
the largest connected subgraph is expected to become more stable as
$m$ increases. If there is an $m$ above which $\aS=\cbm+1$ for
arbitrary large $t$ above some $t_0$ is an open question. Comparing
Fig.~\ref{lECA}(a) and Fig.~\ref{lECA}(b) shows that random attachment
and preferential attachment  have similar $m$-dependence
behavior---the major difference being that preferential attachment has
a much sharper increase of $\aS$; to be more precise the $m$ values
that does not reach $S=\cbm+1$ for any $t$, have a lower value in the
large-$t$ limit.

\begin{figure}
  \centering{\resizebox*{8cm}{!}{\includegraphics{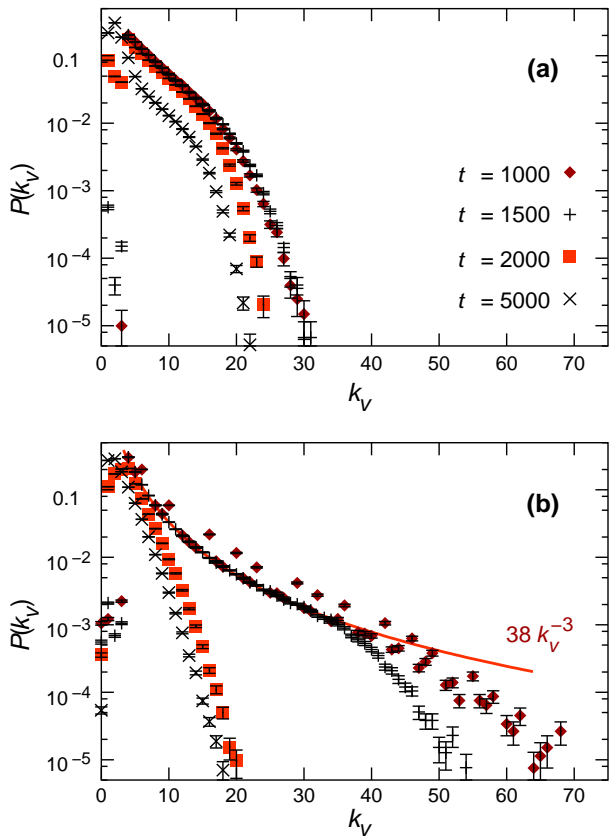}}}
  \caption{Histograms (averaged over $10^4$ runs) of degree. The parameter
    values are $m=m_0/2=4$ and $\cbm=500$. (a) shows histograms for random
  attachment, (b) shows histograms for random attachment. The grey
  line in (b) is the function $38\,k_v^{-3}$ illustrating the
  emerging power-law degree distribution at early times.
  }
  \label{ECAhistK}
\end{figure}

\begin{figure}
  \centering{\resizebox*{8cm}{!}{\includegraphics{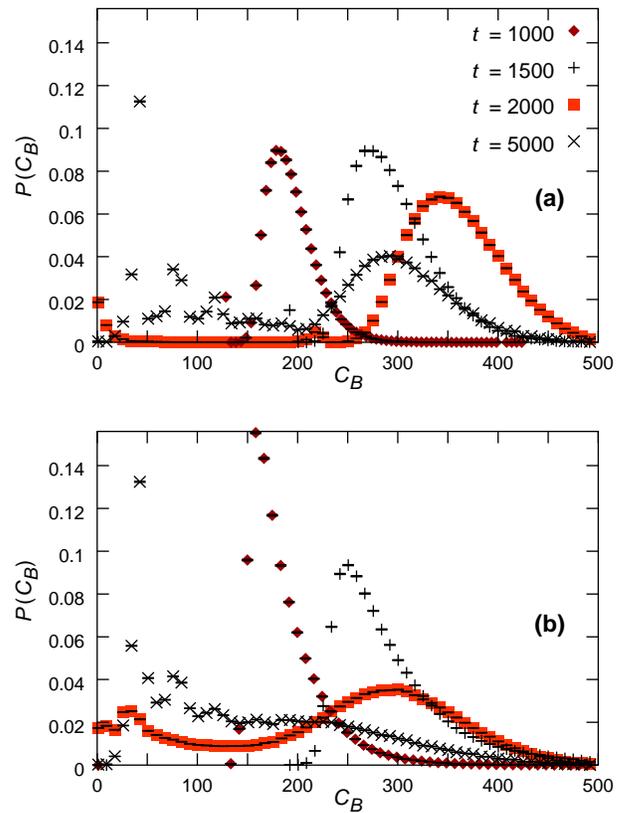}}}
  \caption{Histograms (averaged over $10^4$ runs) of edge
    betweenness centrality. The parameter values are---as in
  Fig,~\ref{ECAhistK}---$m=m_0/2=4$ and $\cbm=500$. (a) shows
  histograms for random attachment, (b) shows histograms for random
  attachment.
  }
  \label{ECAhistB}
\end{figure}

\begin{figure}
  \centering{\resizebox*{8cm}{!}{\includegraphics{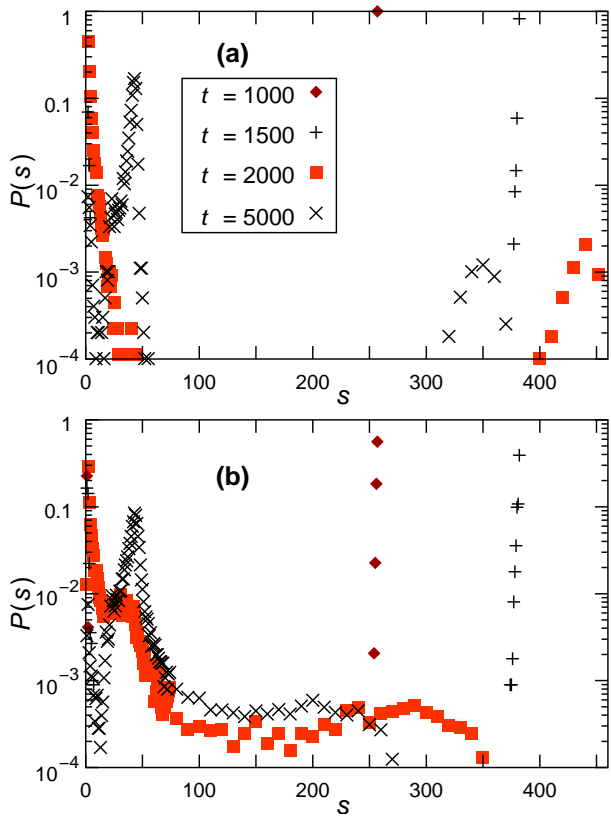}}}
  \caption{
    The distribution of sizes of connected clusters at different times
    for $m=m_0/2=4$, $\cbm=500$ and (a) random attachment, (b)
    preferential attachment. All curves represents averages over
    $10^4$ runs. To overcome noise the $t\geq 2000$ histograms are
    binned for $s\geq 75$, each point being an average over a width
    of 10.
  }
  \label{ECAhist2}
\end{figure}

To get another angle of the mechanisms of the breakdowns for small $m$, we
consider histograms of degree $k_v$ and betweenness
$C_B(e)$. Figs.~\ref{ECAhistK} and \ref{ECAhistB} shows these
histograms both before and after the large drop in $\aS$ for
$m=m_0/2=4$ and $\cbm=500$. (In the random attachment case this drop
occurs at $t_\mathrm{drop}\approx 1600$, the corresponding value for
preferential attachment is  $t_\mathrm{drop}\approx 2000$.) For random
attachment the difference between the
histograms before and after the $\aS$-drop is distinctively smaller
than for preferential attachment, just as expected from
Fig.~\ref{lECA}. The random attachment curves in Fig.~\ref{ECAhistK}(a)
has a degree distribution of truncated exponential form both
at the earlier and later times. In Fig.~\ref{ECAhistK}(a) it is exponential
over two decades of $P(k_v)$, but falls off faster than exponentially
for higher $k_v$. For preferential attachment the degree distributions
(Fig.~\ref{ECAhistK}(b)) have a distinct difference---at
$t<t_\mathrm{drop}$ there is an emergent power-law shape of the
$P(k_v)$-curve, whereas at $t>t_\mathrm{drop}$ the shape is
exponential, $\sim\,\exp(-0.62\, k_v)$, over five decades. To
summarize, the degree distributions before and after the $\aS$-peak
illustrates the same behavior as the time evolution of $\aS$---the
breakdown in the preferential attachment case is both faster and more
restructuring than in the random attachment case.

The betweenness distributions of Fig.~\ref{ECAhistB} shows a peak that
moves to higher $C_B$, as $t$ grows, until it reaches its maximal
value at the time of the drop in $\aS$ and starts to decrease. For
random attachment (Fig.~\ref{ECAhistB} (a)) the shape of the
distribution looks qualitatively the same before and after the drop,
but for preferential attachment (Fig.~\ref{ECAhistB} (b))
$P(C_B)\approx 0$ for betweenness smaller than the peak. The vertex
betweenness distribution of the BA model is known to be strictly
decreasing~\cite{KAHNG}, which would imply that the low-$C_B$ tails in
Fig.~\ref{ECAhistB}~(b) (and most likely in Fig.~\ref{ECAhistB}~(a) as
well) comes from a spread of the size of the largest cluster, rather
than from a tail in the largest cluster's betweenness
distribution. Another feature of the betweenness histograms of
Fig.~\ref{ECAhistB} is the smaller peaks at low $C_B$ for
$t<t_\mathrm{drop}$. These peaks corresponds to a sharp peak of the
cluster size distribution just after the $\aS$-peak (see
Fig.~\ref{ECAhist2}). Such smaller clusters have small average degree
with many $k_v=1$ vertices, which all contributes to a peak at $s$ of
the betweenness histograms. This explains the peak at $C_B\approx 45$
in the $t=5000$ curve of Fig.~\ref{ECAhistB}~(b).

The distribution of cluster sizes displayed in Fig.~\ref{ECAhist2}
gives some further insights: For $t>t_\mathrm{drop}$ of the random
attachment curves shows a bimodal distribution as $P(s)$ is zero in
the interval $60\lesssim s\lesssim 290$. The preferential attachment
curves, in contrast, has a long tail. Both the large $s$ peak for
random attachment and the tail of preferential attachment corresponds
to one single cluster. This is in striking contrast to the vertex
overload case~\cite{EGO2} where the network looses the unique largest
component after the breakdown avalanches. As $t$ evolves well beyond
$t_\mathrm{drop}$ the largest component peak decreases, and does thus
not represent a giant component (a largest cluster proportional to
$N$). The picture for both random and preferential attachment is thus
that the system does not loose its unique largest cluster in a single
breakdown avalanche---an avalanche rather results in a few isolated
vertices or smaller clusters getting disconnected from the largest
connected component.

The overall picture of the time evolution of $\aS$, $L$ and $\il$
(Fig.~\ref{ECAcase}), the $m$-dependence (Fig.~\ref{lECA}), as well as
the histograms of Figs.~\ref{ECAhistK}, \ref{ECAhistB} and
\ref{ECAhist2} is that for small $m$, avalanching breakdowns fragments
the network to a state from which it never recovers. For preferential
attachment the newly fragmented network contains a single largest
cluster with a well defined size, and the emergent scale-free degree
distribution before $t_\mathrm{drop}$ is replaced by an exponential
distribution. The breakdown for the random attachment case turns out
to be less violent, and does not cause any major structural
change. Furthermore, the difference between the random and
preferential attachment cases is larger for edge breakdown than for
the corresponding vertex breakdown model studied in Ref.~\cite{EGO2}.

\section{Summary and conclusions}
We have studied networks grown by the Barab\'{a}si-Albert model for
networks with emergent scale-freeness and edges sensitive to
overloading. Except the preferential attachment defining the BA
model, we also study an unbiased random attachment. We focus on the
case where the number of established connections to random other
vertices of the network scales linearly with the number of vertices in
the network.

We find that for intermediate values of $m$ (the number of edges added
per vertex) the network grows like the BA model up to a point where is
starts to break down. After a number of avalanching breakdowns the
network reaches a state characterized by many disconnected clusters
from which a giant component never re-emerges (although, in the
preferential attachment case, there will always be one single largest
cluster much larger than any other). If the growth is by random
attachment, the breakdown is less violent with smaller avalanches and
no pronounced structural change. For large $m$ the steady state at
large times is characterized by a constant largest cluster size.

In context of real world communication networks one can conclude that
these would benefit from being grown by random rather than
preferential attachment (and this difference being larger for edge
overload than for vertex overload studied in Ref.~\cite{EGO2}). In the
vertex overload case avalanches proceeds until the network is
fragmented into small clusters; in the edge overload problem there is
still one large component after the breakdowns, thus we infer that for
real-world communication networks, vertex overloading is a greater
threat than edge overloading, and congestion control in telecommunication
networks~\cite{kihl} and Internet routing protocols~\cite{huitema}
should focus on balancing the vertex rather than edge load. Only if
the capacity of vertices (servers etc.) grows significantly faster
than the capacity for edges (cables etc.), edge overload breakdown is
a potential threat for avalanching breakdowns that is triggered by the
change of load in a growing network.

\section*{Acknowledgements}
The author thanks Beom Jun Kim for discussions. This work was
partially supported by the Swedish Natural Research Council through
Contract No.\ F~5102-659/2001.

\appendix

\section{Intrinsic communication activity}\label{sec:app}

This paper deals mainly with the case where the average user of a
growing communication network communicates with a number of
others that increases linearly with $N$. One can also imagine a case
where, even though the network grows, the users in average
communicates with a network size independent number of others; which is
the topic of the present Appendix. (In Ref.~\cite{EGO2} this scenario
was termed ``intrinsic communication activity.'') The behavior of real
communication networks lies, presumably, between these two extremes.

\begin{figure}
  \centering{\resizebox*{8.6cm}{!}{\includegraphics{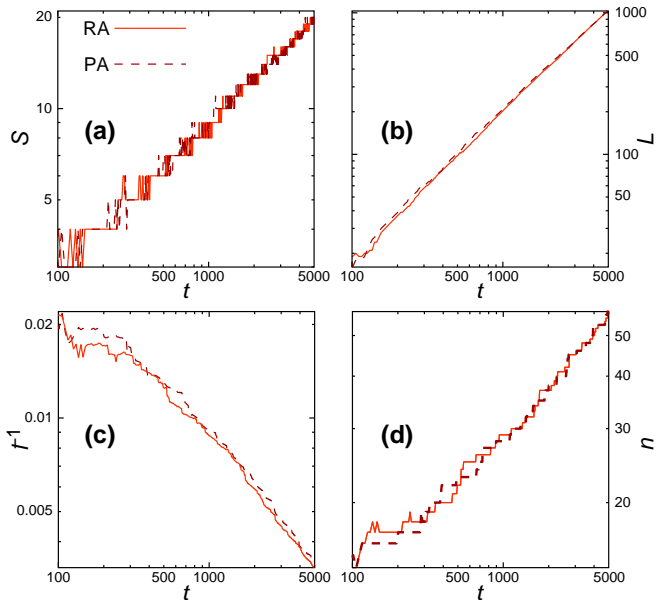}}}
  \caption{The time evolution of $S$ (a), $L$ (b), $\il$ (c), and $n$
    (d) for a typical run in the intrinsic communication activity
    case. The model parameters are: $\ibm= 0.1$ and
    $m_0/2=m=5$. Dashed lines represent the network grown with
    preferential attachment (PA), solid grey lines denotes curves for
    the runs with an unbiased random attachment (RA).
    }
  \label{ICAcase}
\end{figure}

\subsection{Definitions}

To implement the situation of intrinsic communication activity, we
modify Eq.~\ref{eq:load_eca} to:
\begin{equation}
  \Omega'=\{\Lambda:|\Lambda|=A'N\}
\end{equation}
where $A'$ is constant with respect to $N$. I.e.\ the users has the
$N$-independent average number $A'$ of established contacts through
shortest routes. Averaging the load over $\Omega'$ according to
Eq.~\ref{eq:lambda} gives:
\begin{eqnarray}
  \langle\lambda(e)\rangle_{\Omega'}&=&\frac{1}{|\Omega'|}
  \sum_{\Lambda\in\Omega'} \sum_{(w,w')\in\Lambda}
  \frac{\sigma_{w\,w'}(e)}{\sigma_{w\,w'}} \nonumber \\
  &=& \frac{A'}{N}\,C_B(e) ~.
\end{eqnarray}
From this we see that having a constant capacity for the load
$\lambda(e)$ corresponds to having a limit on $C_B(e)$ that increases
with $N$. Thus we view $e$ as overloaded if $C_B(e)$ exceeds
$\cbm=N\ibm$ (where $\ibm$ is constant).

\subsection{Results}

In the vertex overload breakdown problem, the case of intrinsic
communication activity has a more complex dynamics than the extrinsic
communication activity case (studied in the main part of the text),
with giant components forming only occasionally for some sets of
parameter values~\cite{EGO2}. For edge overload breakdown, on the
other hand, the dynamics of a system with intrinsic communication
activity seems very simple with no avalanching breakdowns and no
qualitative difference between preferential and random attachment, see
Fig.~\ref{ICAcase}. We can also notice that the measured quantities
has a power-law dependence of $t$. (Fig.~\ref{ICAcase} is constructed
from one run with random and preferential attachment respectively.)
For large times ($1000\lesssim t\lesssim 5000$) the exponent $\alpha$
for the time development of the respective quantity is (in the large
$t$ limit): $\alpha_{\il}\approx -0.6$, $\alpha_L=1.0$, and
$\alpha_S=\alpha_n=0.50$ for both (a) and (b). Initially
$\alpha_{\il}$ is closer to zero, for $100\lesssim t\lesssim 1000$ we
have $\alpha_{\il}\approx -0.5$. To illustrate the consistency of the
exponents we note that
\begin{equation}
  \sum_{e\in E} C_B(e)=\sum_{v\in V}\sum_{w\in V\setminus\{v\}}
  d(v,w)=n\frac{N}{n}\left(\frac{N}{n}-1\right)
  \langle l_\mathrm{CS}\rangle~,
\end{equation}
where $\langle l_\mathrm{CS}\rangle$ is the average geodesic length
for a connected subgraph, and $d(v,w)=0$ if $v$ and $w$ are
disconnected. This yields
\begin{equation}\label{eq:CBv}
\langle C_B(e)\rangle \approx m\left(\frac{N}{n}-1\right)
l_\mathrm{CS} \leq \max_{v\in V}C_B(e) ~.
\end{equation}
If one assumes that $\langle C_B(e)\rangle\propto\max_{v\in V}C_B(e)$
and $N\gg n$ we have $\langle l_\mathrm{CS}\rangle\propto n$. Making
the crude approximation $\il\approx\langle l_\mathrm{CS}\rangle^{-1}$
gives $\alpha_{\il}\approx-\alpha_n$, which holds well for small
$t$. As $t$ increases the spread in shape of the connected subgraphs
becomes larger so the $\il\approx\langle l_\mathrm{CS}\rangle^{-1}$
approximation becomes worse which is seen as a slight increase of the
slope $\alpha_{\il}$. That the approximation
$\alpha_{\il}\approx-\alpha_n$ is rather good throughout the range of
$t$ is also reflected in that the average size of connected components
$N/n$ is never very far from $S$: At $t=5000$ we have (see
Fig.~\ref{ICAcase}(a)) $N/n\approx 50$ and $S=57$ for random
attachment, and $N/n\approx 53$ and $S=56$ for preferential
attachment. In this approximation we see that $l\propto n\propto
N^{1/2}$ so the small average geodesic length is lost within the
connected subgraphs.
If $\ibm$ is chosen larger, so the network initially grows without
edges breaking, there are no large avalanches but a crossover to the
behavior seen in Fig.~\ref{ICAcase}.

\end{document}